\documentstyle[twocolumn,aps,graphicx]{revtex}

\begin{document}

\hyphenation{Ka-pi-tul-nik}

\twocolumn[
\hsize\textwidth\columnwidth\hsize\csname@twocolumnfalse\endcsname
\draft

\title{Inherent Inhomogeneities in Tunneling Spectra of BSCCO Crystals in the
Superconducting State}

\author{C. Howald$^1$, P. Fournier$^2$,  and A. Kapitulnik$^1$ }

\address{$^1$Departments of Applied Physics and of Physics, Stanford
University, Stanford, CA 94305, USA\\
$^2$Center for Superconductivity Research, Department of Physics,
University of Maryland, College Park, Maryland 20742, USA\\
}
\date{\today}

\maketitle

\begin{abstract}
Scanning Tunneling Spectroscopy on cleaved BSCCO(2212)
single crystals reveal inhomogeneities on length-scales of $\sim$30
$\AA$.  While most of the surface yields spectra consistent with a
d-wave  superconductor, small regions show a doubly gapped
structure with both gaps lacking coherence peaks and the larger
gap having a size typical of the respective pseudo-gap for the
same sample.
\end{abstract}

\pacs{PACS numbers: 74.72.Hs, 74.50.+r, 74.25.-q }
]


Tunneling spectroscopy has been an important tool in the study of
high-temperature superconductors since their  discovery. While in
the  early days of high-Tc a variety of gap sizes  and structures
were found and introduced much  controversy into the subject,
later measurements yielded greater consistency among groups,
revealing a more  coherent picture of the surface of high-Tc
materials as viewed with  STM.  Among many examples we note that
STM studies revealed the  nature of the  superstructure in BSCCO
\cite{kirk}, the d-wave nature  of the gap and  its size \cite{renner1},
the effect of local impurities  and the  emergence of zero-bias anomalies
\cite{yazdani,davis1,davis2}  , and the  electronic structure of the
core of vortices  \cite{renner2,davis3}.

While in general published data concentrates on specific aspects
of the properties of the superconductor such as the gap, the
pseudo-gap, impurities, etc., very little has been published on
the large-scale structure of the electronic state at the surface
of high-Tc materials. Judging from photoemission measurements
that reveal a $\vec{k}$-dependent gap and pseudo-gap
\cite{shen1,shen2,ding,loeser}, thus  asserting that $\vec{k}$ is
a good quantum number, it has been the  common belief that the
surface is homogeneous with only few scattering  centers and that
it represents the bulk properties. However, recent measurements
by a variety of techniques  suggest that superconductivity may
not be homogeneous in high-Tc superconductors. In particular,
STM measurements of cold-cleaved YBCO
have found that the gap along the chains is inhomogeneous and correlates with
oxygen vacancies \cite{lozanne}. The issue of homogeneity  (either in the
static or dynamical sense) is very important in view  of recent theoretical
developments that find phase separation to be an  integral part of
the high-Tc scenario \cite{ek1,ek2,ek3,ek4,laughlin}.

In this paper we discuss scanning tunneling spectroscopy  measurements
on the surface of $\rm Bi_2Sr_2 $ $\rm CaCu_2O_{8-x}$  (BSCCO) single
crystals which show that superconductivity in high  temperature
superconductors is inherently inhomogeneous. This inhomogeneity appears at a
scale of a few coherence lengths and is not a consequence of local impurities
or strong disorder. We argue that this inhomogeneity is a consequence of
electronic phase separation into regions that are proximity-coupled to give a
continuous variation of the tunneling spectroscopic features.

The slightly underdoped single crystals of BSCCO used in this study were grown
by a directional solidification method \cite{fournier} and their
T$_c$ of $\sim$ 80 K was measured by SQUID magnetometry. Crystals were
mounted into a UHV-Low-T STM system. The samples are cleaved at room
temperature in a vacuum of better than 10$^-$$^9$  torr.

\begin{figure}
\includegraphics[width=0.95 \columnwidth]{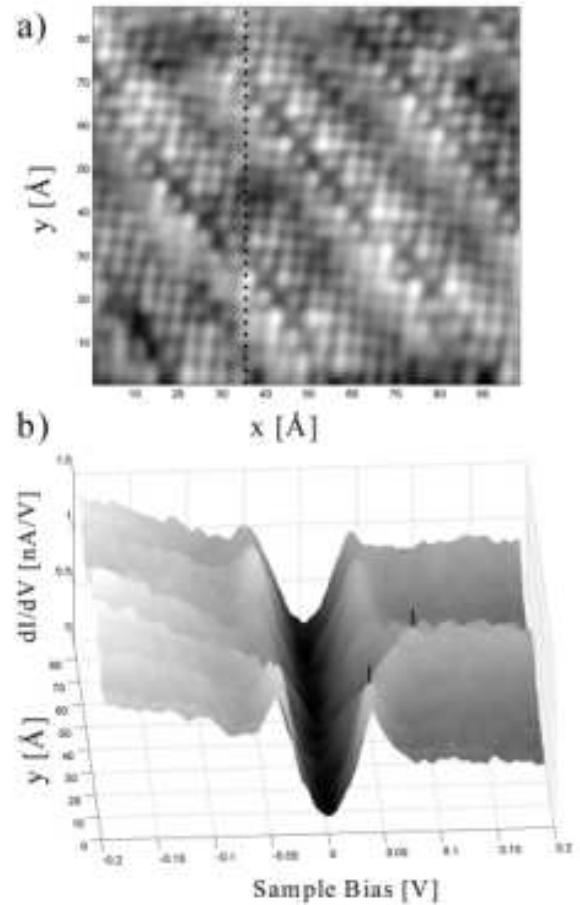}
\caption{Typical topographic scan of a 100 $\AA$ x 90 $\AA$ area
(a). Dotted line marks the location of the spectroscopic scans
shown in the lower figure (b). The two bold lines in the scan
mark the extreme spectra discussed in the text.} \label{fig1}
\end{figure}

\noindent  Immediately, they are
transferred to the low temperature chamber
which houses the STM and presumably
reaches much lower base pressure because the walls of this
portion of the chamber are held at liquid helium temperature. The
base temperature of  the microscope is 6 K, however the data
shown is this paper were taken at 8 K. Images are taken with a
gold tip at a sample bias of -200 mV and a setpoint current of
100 pA. The dI/dV spectra are taken using a lock-in technique
with an AC modulation of 1 mV, after fixing the tip location at
a  height that gives -100 pA current at -200 mV. This setpoint
current establishes the relatively arbitrary normalization condition
used for the raw data shown in this paper. The
quality of the surface and the scanning capability are demonstrated in
figure 1a where we show a topographic scan of a 100 $\AA$ x 90
$\AA$ region. Evident are the atomic resolution and the clarity
of the superstructure \cite{kirk} with average periodicity $\sim$ 27
$\AA$. All cleaves which yield images exhibit similar topography.

Figure 1b shows a line of differential conductance spectra taken
over a range of 80 $\AA$ in the region shown in figure 1a. By
fitting the acquired spectra to a d-wave density of states for a
maximum gap size $\Delta$ ($\Delta (\theta)=\Delta cos(2\theta)$) with
gaussian broadening $\Gamma$
\cite{oda}:

\begin{equation}
{{N_S(E)}\over{N_N}}=\int_{-\infty}^{\infty}Re{\int}_{0}^{2\pi}{{{d\theta}\over{2\pi}}{{E}\over{\sqrt{E^2-\Delta(\theta)^2}}}
{{e^{-{{(E-U)^2}\over{2\Gamma^2}}}\over{\sqrt{2\pi}\Gamma}}}dU,}
\end{equation}

\noindent we find
two important results. For most of the spectra this fit yields
values of $\Delta=42\pm$2 mV and $\Gamma=4\pm$2 mV with residuals
approximately the size of the noise.  However, there are regions
in which the spectra give poor fits to this formula, also
yielding unphysically large values of $\Gamma$ and $\Delta$. The
regions are 20 to 30 $\AA$ across, their areal fraction is $\sim$ 20$\%$, and
they seem to be centered along the superstructure  ridges. The size and shape
of one of the regions is shown in figure 2a, which maps the gap size
(normalized by the 42 mV value found above) as determined by the
d-wave fit. This map, like figure 1b, shows that the spectra vary
continuously through each region.

This continuous variation suggests two extremal spectra: those
from the center of these regions and those far away. These two
extreme shapes are shown in figure 3 for positive bias. In
particular, the spectrum taken outside the regions are consistent
with the density of states of a d-wave superconductor averaged
over k-space (as expected for tunneling perpendicular to the Cu-O
planes), while the other shows a double gap shape with no
coherence peaks.  In addition, the shape of the smaller of these
two gaps is similar to the shape of the inner part of the
superconducting spectra. Figure 2b shows the width at which the
conductance reaches twice the zero bias value, normalized by the
value of those spectra with the best d-wave fits. This plot uses
the same scale as figure 2a in order to emphasize the contrast
between them. The only features in figure 2b are consistent with
noise. That is, while the behavior at the gap energy is
inhomogeneous, at low energies the sample appears to be
homogeneous. The fact that the spectra in these regions look like
those of the background superconductor at low bias indicates that
the node structure exists throughout the sample. Thus, the regions
must maintain the d-wave structure despite the apparent
suppression of superconductivity. In particular, this rules out
these features being caused by impurities. Both theoretical
\cite{balatsky,kashiwaya} and experimental studies
\cite{yazdani,davis1,davis2} show that potential scatterers yield
entirely different behavior, characterized by subgap structure
(zero-bias anomalies) and no change in the value of the gap.

\begin{figure}
\includegraphics[width=1.0 \columnwidth]{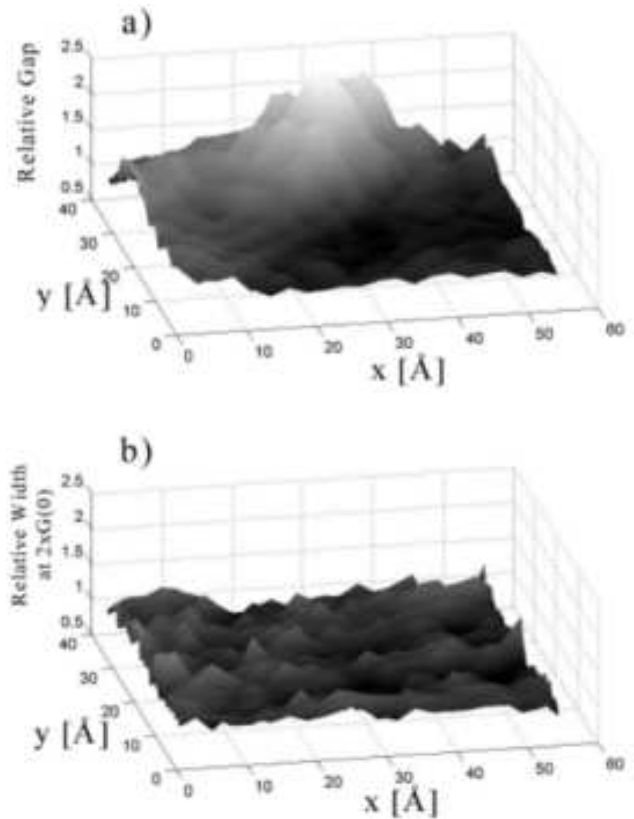}
\caption{Map of normalized gap size (a). High points reflect a
larger gap with suppressed coherence peaks. Lower map (b) is the
width of the spectrum at twice the minimum conductance G(0), indicating
a homogeneous gap structure at low energies (see text). }
\label{fig2}
\end{figure}

Examination of figure 1 and other similar scans shows that the
two extreme shapes are separated by a distance of $\sim$ 30
$\AA$. For BSSCO this distance is roughly twice the coherence
length $\xi$. This suggests that the smooth variation seen may
not be the product of smooth variation in the underlying
character of the material, but may instead reflect a breakdown
into domains with different character which are then proximity
coupled. This proximity coupling would explain why these regions
do not break the superconductors d-wave symmetry.

\begin{figure}
\includegraphics[width=0.95 \columnwidth]{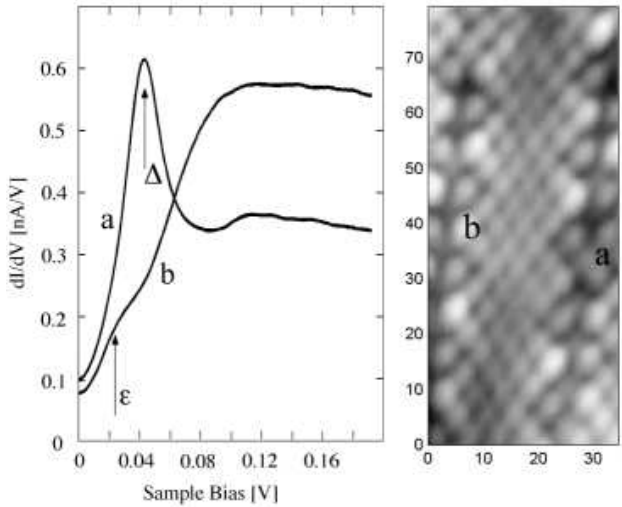}
\caption{Extremal dI/dV from the region shown in figure 2.
Location of the spectra indicated in the topograph (right). Note that the two
spectra are taken at equivalent positions with respect to the superstructure.
In the spectra, $\Delta$ represents the gap energy while $\epsilon$ marks the
kink energy, related to the size of the ``bad" superconductor region (see
text).}
\label{fig3}
\end{figure}

Although curve (b) in figure 3 has the same shape at low bias as
a ``good" superconductor, it develops a different gap structure at
higher bias. The abrupt change in gap structure is an indication
that the spectrum of quasiparticles changes abruptly. HTSC, and
in particular BSCCO crystals of the type we are using are
believed to be in the clean limit. The superconducting state is
therefore characterized by a gap $\Delta$, associated with
pairing instability and ``pair size" $\xi \sim \hbar v_F /
\Delta$. Terminating the gap opening at energies $\epsilon <
\Delta$, means a larger length scale $L \sim \hbar
v_F/\epsilon$.  Since the coherence peaks mark roughly the
coherence length, the shoulder in curve (b) indicates a length
scale that is roughly twice the coherence length. This is therefore the
length at which the system crosses over from the behavior of a pure d-wave
superconductor at long length scales to behavior which is more
reminiscent of the ``insulating" state. The larger gap with no
coherence peaks resembles the pseudo-gap.

Summarizing our observations, we find that while the gap changes
smoothly between the two extremes presented in figure 3, the
system breaks into distinct domains of either ``good" or ``bad"
superconductivity. This observation suggests that the ``bad"
superconducting regions are a consequence of insulating regions
in proximity with the ``good" superconductor. Since for high-Tc
superconductors the coherence length is so short, proximity
implies leakage of the superconducting wavefunction a distance
of  order $\xi$ into the insulator\cite{deutscher}. Similarly,
the superconductor weakens as well over a similar distance close
to the boundary. The result is therefore the smoothly modulated
behavior shown in figure 2. This follows from a scenario for
phase separation where the initial boundaries between adjacent
puddles are relatively sharp. Such a situation is most likely a
consequence of a proximity to a first order phase transition
that  is becoming continuous due to disorder
\cite{ek1,ek2,ek3,ek4,laughlin,sak} as was first shown by Imry
and Ma \cite{imry}. The sharpness of the initial domains is
indicated by the relatively sharp kink in the spectrum of the
``bad" superconductor as it crosses-over from low energy
superconductor-like to high energy pseudo-gap-like behavior.

The above suggestion that the inhomogeneities observed by STM
spectroscopy on the surface of BSCCO single crystals are intrinsic
and reflect phase separation does not contradict the photoemission
results \cite{shen1,shen2,ding,loeser}. While one might expect
that inhomogeneities would destroy the k-dependence observed in
photoemission, the type of inhomogeneities seen here would not.
The low-k behavior would measure the long-wavelength homogeneity,
while, since both the superconducting gap and the pseudo-gap have
the same d-wave symmetry, averaging in regions of pseudo-gap would
not destroy the symmetry of the gap seen in photoemission. Moreover,
the above scenario is also in agreement with recent
investigations of vortex structure in High-Tc superconductors.
The peculiar result there is the fact that while the zero-field
spectroscopy maps show an inhomogeneous pattern, application of a
magnetic field results in vortices that are always surrounded by
``good" superconductor as judged from the STM spectroscopy
\cite{renner2,davis3}. Within the framework of our interpretation
this implies that vortices will prefer to have their cores reside
on ``bad" superconducting regions. The surrounding area is then a
good superconductor and since the vortex core does not proximity
couple to the superconductor, the result is a homogeneous
superconductor in the regions surrounding the vortex.

Having argued that the observed inhomogeneities are not the
consequence of disorder, we can check this claim by intentionally
disordering the surface. Since BSCCO is very sensitive to
excessive current density or electric field we locally increased
the setpoint current to 500 pA at -200 mV sample bias and scanned
the tip over  a 140 $\AA$ x 90 $\AA$ area. Subsequent topography
shows an apparent depression at the center of the ``damaged" area
because of decreased local density of states. The superstructure
and atomic corrugation are no longer visible within this region,
indicating strong disorder, though they remain visible over the
rest of the sample surface. Figure 4 shows a strip-image of the
contour of constant density of states and a line of spectra taken
across the ``damaged" area. Clearly at about 100$\AA$ away from
the center of that area a good superconducting gap is obtained
indicating only very small influence of the damaged area. As we
approach the  boundary of the scanned area and continue towards
its center the  spectra change continuously. In particular, the height of the
coherence peak drops continuously to the background conductance level.
However, unlike the variations in  the spectra shown in figure 1, here the
gap size does not change  at all as the coherence peaks disappear. Once
inside the region, the gap broadens into a pseudo-gap, and then an insulating
gap. The last two spectra that are inside the damaged area look very similar
to those of the pseudo-gap previously observed by tunneling \cite{renner2}
and photoemission \cite{loeser}.

\begin{figure}
\includegraphics[width=0.95 \columnwidth]{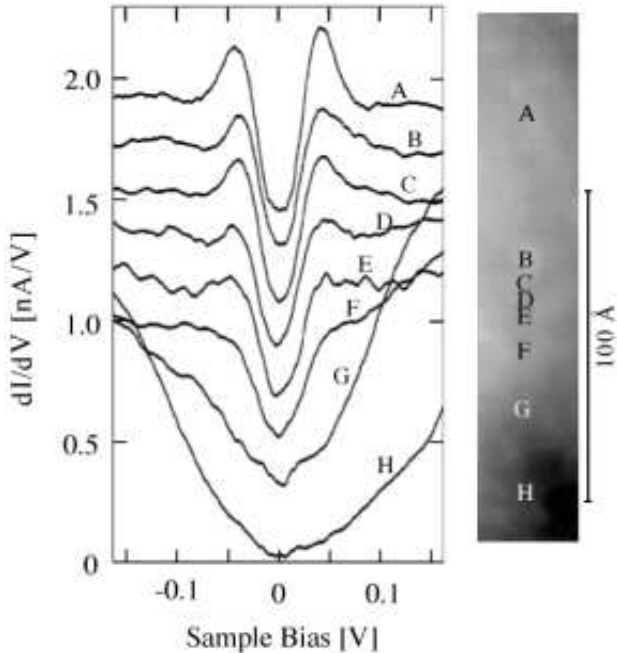}
\caption{Spectra taken near the disordered area. The location of
each spectra is shown in the topograph (right).} \label{fig4}
\end{figure}

Comparing this area with the region on the native surface,
several points are obvious. Both show a continuous decrease in the
coherence peak height. Also, both show pseudo-gap features inside
the region. However, the pseudo-gap size varies in the
intentionally disordered case while it seems to be constant in
the other. This is presumably because the disorder, and perhaps
doping as well, is varying throughout this region, yielding a
varying intrinsic gap size. Secondly, for the native defect,
there remains a vestige of the superconducting gap in the
pseudo-gapped spectra. This indicates that the smaller size of
the native defect allows proximity coupling to the superconductor
that is lost inside the larger manufactured defect.  The effect
of disorder is therefore different than the intrinsic
inhomogeneities observed on the pure surface. When strong
enough,  the disorder ``pins" the insulating phase as is the case
in the  middle of the damaged area. Again, this implies a
``good"  superconductor at the periphery of the ``bad"
superconductor.

In conclusion, we find that the intrinsic surface of BSCCO
exhibits inhomogeneities that strongly suppress
superconductivity, while maintaining the same low energy
structure. These regions show spectra that are reminiscent of the
pseudo-gap, but also are proximity coupled to the surrounding
superconductor.


We thank Steve Kivelson and Bob Laughlin for many useful discussions.
Work supported by Air Force Office of Scientific Research.

\end{document}